\documentstyle[12pt]{article} 
\begin{document}
{\obeylines
June 1996 \hfill ULB-TH-96/07\\[-5mm]
\hfill  PAR--LPTHE 96-27\\[-5mm]
\flushright hep-th/9606179
}
\vskip 2 cm

\begin{center}

 {\bf THERMODYNAMICS OF BLACK HOLES IN PRESENCE OF STRING
INSTANTONS\footnote{Presented by F. Englert at the Second International
Sakharov Conference on Physics, Moscow, May 20-24, 1996.}}\\ \vskip 2 cm

{\bf Fran\c cois Englert$^a$, Laurent Houart$^b$\footnote{Chercheur
associ\'e CNRS and Charg\'e de Recherches FNRS \`a titre honorifique.}
 and Paul Windey$^b$}\\  
\bigskip

{\it a) Service de Physique Th\'eorique,\\ Universit\'e Libre de Bruxelles, 
Campus Plaine, C.P.225,\\ Boulevard du Triomphe, B-1050 Bruxelles, Belgium\\ 
and\\
 b) Laboratoire de Physique Th\'eorique et Hautes Energies\\  
Universit\'e Pierre et Marie Curie,\\ Paris VI, Bte 126, 4 Place Jussieu,
75252 Paris cedex 05, France\footnote{Laboratoire associ\'e No. 280 au CNRS.}
}
\end{center}
\bigskip\bigskip\bigskip\bigskip
\noindent
{\bf Abstract.}\quad The coupling of a Nambu-Goto string to gravity allows for
Schwarzschild black holes whose entropy to area relation is
$S=(A/4)(1-4\mu)$, where $\mu$ is the string tension. The departure from $A/4$
universality results from a string instanton which leads to a
materialisation of the horizon at the quantum level.
\vfill \eject
\section{Thermodynamics}
 
The possibility of departing from $A/4$ universality can be understood in
thermodynamic terms. Consider the differential mass formula~\cite{bch} for
black holes surrounded by static matter in the form~\cite{ce}   
\begin{equation} 
\label{dmf}
\delta M_{tot}={\kappa\over 2\pi}{\delta A\over 4} + 
\sum_i\partial_{\lambda_i}H_{matter}\delta\lambda_i,
\end{equation} 
where $M_{tot}$ is the total mass, $\kappa$   the surface gravity of the hole,
$A$ the area of the event horizon and the $\lambda_i$ are all the parameters in
the matter action. It is important to realize that the derivation of this
identity involves only classical physics and cannot involve the Planck constant.
If one admits the Bekenstein assumption~\cite{bek} that the black hole
contributes to the entropy one can interpret Eq. (\ref{dmf}) as the expression
of the first principle of thermodynamics,   $\partial_{\lambda_i}H_{matter}$
being the generalized forces, by identifying $A/4$ to the entropy up to a
multiplicative constant proportional to $\hbar^{-1}$. The temperature will
then be proportional to $\hbar$. 

It is well known that to compute thermal correlation functions and
partition functions in field theory in flat Minkowski spacetime one can
use path integrals in periodic imaginary time.  The period $\beta$ is the
inverse temperature and can be chosen freely. This method was generalized  to
compute matter correlation functions in static curved backgrounds.  For the
Schwarzschild black hole, possibly surrounded by matter, the analytic
continuation to imaginary time defines a Euclidean background everywhere except
at the analytic continuation of the horizon, namely the 2-sphere at $r=2M$.  
Gibbons and Hawking~\cite{gh1} extended the analytic continuation to the
gravitational action, restricting the hitherto ill-defined path integral over
metrics to a saddle point in the Euclidean section.   To constitute such a
saddle the Euclidean black hole must be regular given that a singularity at
$r=2M$ would invalidate the solution of the Euclidean Einstein equations. This
implies a unique   Hawking temperature $T_H$ which, in natural units, is always
equal to $\kappa / (2\pi)$.   Thus  Eq. (\ref {dmf}) yields the
Bekenstein-Hawking area entropy $S$ for the black hole    
\begin{equation} 
\label{area}
 S = {A\over 4}.   
\end{equation} 
The entropy  Eq. (\ref {area}) is not affected by mass surrounding
the black hole and would therefore  seem   to depend only on
the black hole mass. This is not  the case: a different relation between entropy
and area arises when a conical singularity is generated in the
Euclidean section~\cite{ehw1} at $r=2M$.
 
As pointed out by many authors,    a conical singularity at
$r=2M$ modifies the Euclidean periodicity of the black hole and hence its
temperature. If  this singularity would arise from a source term in the
Euclidean Einstein equations,  Eq. (\ref {dmf})  would remain valid and could be
written as  
\begin{equation} 
\label{dmfnew}
\delta M_{tot}=T\delta [(1-\eta) {A\over 4}] + T{A\over 4}\delta \eta +
\sum_i\partial_{\lambda_i}H_{matter}\delta\lambda_i,
\end{equation} 
where $\eta$ is the deficit angle and the temperature $T$ is related to the
Hawking value $T_H$ by $T=T_H (1-\eta)^{-1}$. It follows from  Eq. (\ref{dmfnew})
that a new generalized force $X_{\eta}$, conjugate to $\eta$,  
\begin{equation} 
\label{force}
X_{\eta}=T {A\over 4}
\end{equation}
must appear and that the entropy of the hole would become
\begin{equation} 
\label{areanew}
 S =(1-\eta) {A\over 4},   
\end{equation} 
independent  of the   surrounding matter.

We now show that a deficit angle can be generated by a string instanton and
that Eqs. (\ref {force}) and (\ref {areanew}) obtain with  
 $\eta$ determined by the string tension. 

\section{The String Instanton}

For simplicity we describe here only the case of a pure black hole of mass $M$.
The general case is treated elsewhere~\cite{ehw2}. In presence of a Nambu-Goto
string the Euclidean action is \begin{eqnarray}
\nonumber
  I &=&-{1\over 16\pi}\int_{  M}\ \sqrt{ g }
R +{1\over 8\pi}\int_{\partial   M}\ \sqrt{ h }K \\
\label{action}
&&- {1\over
8\pi}\int_{(\partial   M)_\infty}\ \sqrt{ h_0 } K_0 + \mu \int\
d^2\sigma \sqrt{\gamma}.
\end{eqnarray} 
Here $\mu$ is the string tension and $\gamma$  determinant   of the induced
metric on the world sheet. The latter is taken to have the topology of a
2-sphere. The variation of this action with respect to the metric
  gives the   Einstein equations and the variations with respect to the string
coordinates in   $ \gamma$ give rise to the stationary area condition for the
string.

The Einstein equations  still admit ordinary   black hole solutions
corresponding to zero string area.  The Euclidean space is regular at $r=2M$ and
the $t$-periodicity is the inverse Hawking temperature~\cite{haw} 
\begin{equation}
\label{period} 
\beta_H=8\pi M.
\end{equation}
However there exists a non-trivial solution  to the string
equations of motion in Euclidean space when the string wraps around the
Euclidean continuation of the horizon, a sphere at $r=2M$. This solution has
 a curvature singularity at $r=2M$. Expressing the curvature in
the trace of Einstein equations as  the product of the
horizon times a two dimensional curvature and using the the Gauss-Bonnet theorem
for disc topology  tell us that there is a conical singularity with
deficit angle $ 2 \pi \eta$ such that 
\begin{equation}  
\eta=4 \mu.
\end{equation}
This deficit angle is the sole effect of the string instanton.  It raised the
  temperature from $\beta^{-1}_H$ to $\beta^{-1}=\beta_H^{-1} / (1-4 \mu)$.
 
We now evaluate the free energy of the black hole. The contribution of the
string term to the action Eq. (\ref{action}) exactly cancels the
contribution of the Einstein term. The only
contributions comes from the boundary terms and one gets 
\begin{equation}
\label{free}
F(\beta,\mu) = \beta^{-1}   I_{saddle} = {M \over 2}= {\beta \over 16 \pi
(1-4\mu)}. 
\end{equation}
 From Eq. (\ref{free}) and the thermodynamic relations $S = \beta^2   
(\partial F / \partial \beta)_\mu$, $ X_{4\mu}= (\partial F / \partial
4\mu)_\beta $, one recovers Eqs. (\ref{force}) and (\ref{areanew}) with
$\eta=4\mu$.
 
\section{Horizon Materialization at the Quantum Level}

The  string instanton   at $r=2M$ in Euclidean space does not alter
the classical Lorentzian black hole background  which remains
regular on the horizon.   However dramatic   effects occur  at the
quantum level. To illustrate these we consider the toy model
consisting of the $s$-wave component of a free scalar field
propagating on the Schwarzschild  geometry and we neglect the residual
relativistic potential barrier. This amounts to consider a
2-dimensional scalar field  propagating on the radial subspace of
the 4-geometry. One can then compute   the  expectation value of the
energy-momentum tensor of the scalar field using the trace anomaly~\cite{dfu}
and the boundary conditions defined by the temperature $T=T_H (1-4\mu)^{-1}$.
Using the Kruskal light-like coordinates $(U,V)$ one gets in the vicinity of the
horizons 
\begin{equation}   \left<T _{UU}\right>  =  {\mu
\over 6 \pi U^2}\ {1 - 2 \mu  \over (1-4 \mu)^2};\quad  
\left<T _{VV}\right>  =  {\mu \over 6 \pi V^2}\  {1 - 2
\mu  \over (1-4 \mu)^2}.
\end{equation}
Thus, the string instanton induces, in an inertial frame,   a singularity  in
the  vacuum expectation value of the scalar field energy-momentum tensor  
on the horizon.

In the original derivation of the Hawking radiation from local field
theory~\cite{haw}, the global temperature describing this local equilibrium has
the Hawking value Eq. (\ref{period}) and no singularity   appears on the
  horizon. However if unitarity is to be preserved,  as originally
suggested by 't~Hooft~\cite{thooft}, some kind of materialization appears
unavoidable~\cite{thooft,susshot}. The fact that the non local effect of the
string instanton does both alter the entropy and induce at the quantum level a
materialization of the horizon (which is singular in absence of back reaction)
points towards the possibility of a retrieval of information through non local
effects.
\vfill\eject


\begin{thebibliography}{99}

\bibitem{bch}J.M. Bardeen, B. Carter and S.W. Hawking, {\it Commun. Math. Phys.}
{\bf 31} 161 (1973).  

\bibitem{ce} A.~Casher and F.~Englert, {\it Class. Quantum Grav.} {\bf 10} 
2479 (1993). 
 
\bibitem{bek} J.D.~Bekenstein, {\it Phys. Rev.} {\bf D7} 2333 (1973).  

\bibitem{gh1}G.W.~Gibbons and S.W.~Hawking, {\it Phys. Rev.} {\bf D15} 2752  
(1977).

\bibitem{ehw1} F.~Englert, L.~Houart and P.~Windey, {\it Phys. Lett.} {\bf B372}
29  (1996).

\bibitem{ehw2} F.~Englert, L.~Houart and P.~Windey, {\it Nucl. Phys.} {\bf
B458}  231 (1996).

\bibitem{haw} S.W.~Hawking, {\it Commun. Math. Phys.} {\bf 43} 199 (1975) .
 
\bibitem{dfu}  P.C.W.~ Davies, S.A.~Fulling and W.G.~Unruh, {\it Phys. Rev.} {\bf
D13}  2720 (1976).

\bibitem{thooft} G.~'t~Hooft, {\it  Nucl. Phys.} {\bf B256} 727 (1985). 

\bibitem{susshot} L.~Susskind, L.~Thorlacius and J.~Uglum, {\it Phys. Rev.} {\bf
D48}  3743 (1993)  
\end{thebibliography}
\end{document}